\documentclass{ewass_ss4proc}

\editors{Emeline Bolmont \& Sergi Blanco-Cuaresma}
\editorsurl{www.emelinebolmont.com | www.blancocuaresma.com/s/}
\publisher{Zenodo}
\conference{EWASS Special Session 4 (2017): Star-planet interactions}
\conferencedate{2017}

\title{How uncertainties on stellar atmospheric parameters impact exoplanet studies?}
\author{Sergi Blanco-Cuaresma$^{1,2}$ }

\affiliation{$^{1}$ Observatoire de Gen\`eve, Universit\'e de Gen\`eve, CH-1290 Versoix, Switzerland \\
			 $^{2}$ Harvard-Smithsonian Center for Astrophysics, 60 Garden Street, Cambridge, MA 02138, USA}

\shorttitle{Stellar parameters and exoplanets}
\shortauthors{Sergi Blanco-Cuaresma}

\abs{Exoplanet properties depend on how well the host star is characterized. For instance, the stellar atmospheric parameters (i.e., effective temperature, surface gravity and overall metallicity) are needed to derive the stellar mass and radius by using evolutionary stellar models or empirical calibrations. I studied how the stellar uncertainties impact the determination of planetary characteristics. By using high-resolution spectra and iSpec, I evaluated how discrepancies between two different spectroscopic approaches (the equivalent method and the synthetic spectral fitting technique) can significantly affect the stellar radius estimation, hence, the exoplanet radius and its possible chemical composition.}

\begin{document}

\maketitle

\section{Introduction}

The number of exoplanetary systems is growing at a very high pace\footnote{http://exoplanet.eu/catalog/} thanks to the remarkable improvement of the community in data acquisition and processing. Depending on the technique used to detect and characterize the system, different exoplanet parameters can be derived. The two most common and productive methods are:

\begin{itemize}
    \item Radial velocity: Planets inflict a movement to their host star, making it wobble and leading to variations in the speed with which the star moves toward or away from Earth. Thanks to the Doppler effect, this movement can be detected using spectroscopy. Since the system may not be aligned with our line of sight, the observations cannot accurately determine the mass of the exoplanets (which is needed to distinguish between planets and brown dwarfs), but only provide an estimate of its minimum mass ($M \sin(i)$, where $M$ is the planetary mass and $i$ is the inclination angle between the orbital plane and the line of sight). Although, if planet's spectral lines are visible in the spectrum, the planet radial velocity and inclination can be derived and the composition of the planet studied \citep{2014Natur.513..345B}. The eccentricity of the planet's orbit can be measured directly.
    \item Transit: When the exoplanet crosses our line of sight with the star, it produces a small eclipse that can be observed from earth using photometry. The photometric method can determine the planet's radius. In a first order approximation:

\begin{equation}\label{eq:transit}
    \frac{\Delta F}{F} = \frac{R^{2}_{p}}{R^{2}_{*}}
\end{equation}

        where $F$ is the observed flux, $R_{p}$ the exoplanet radius and $R_{*}$ the stellar radius. Additionally, when the planet is blocked by its star, a secondary eclipse is produced and a direct measurement of the planet's radiation can be obtained.
\end{itemize}

Space-based telescopes mission such as CoRoT \citep{2006ESASP1306...33B}, Kepler \citep{2010Sci...327..977B, 2010ApJ...713L..79K, 2010ApJ...713L..87J} and K2 \citep{2014PASP..126..398H} are capable of detecting a huge number of transiting systems, which then can be confirmed and better characterized with ground-based spectroscopic and photometric observations by surveys such as HAT-Net \citep{2004PASP..116..266B}, WASP \citep{2006PASP..118.1407P}, KELT \citep{2007PASP..119..923P} and HATSouth \citep{2013PASP..125..154B}.

Everyday we have more observations to study many different exoplanetary systems, and in the community there are several tools that can fit the radial velocities and light curves to constrain/derive the system properties. EXOFAST\footnote{http://astroutils.astronomy.ohio-state.edu/exofast/exofast.shtml} \citep{2013PASP..125...83E} and the Exonailer\footnote{https://github.com/nespinoza/exonailer} algorithm \citep{2016ApJ...830...43E} are just two good examples, their source code can be downloaded, studied and used freely.

To fully characterize any system, many of these tools and algorithms require the host star mass and radius. In most cases, these parameters are inferred from the stellar atmospheric parameters (i.e., effective temperature, surface gravity and chemical composition) by using stellar evolutionary codes (e.g., Isochrones python package, \citealt{2015ascl.soft03010M}) or empirical calibrations such as \cite{2010A&ARv..18...67T} which links stellar atmospheric parameters with radii and masses. For instance, using transit data it is possible to derive the radius ratio, thus if the stellar radius is independently known then we can obtain the exoplanet radius.

Unfortunately, it is common to combine stellar atmospheric parameters from different heterogeneous studies. The biases due to non-homogeneous stellar analysis then propagate to the characterization of the exoplanets, affecting the comparison of different exoplanetary systems. In this study, I specifically explored how different spectroscopic techniques and their derived atmospheric parameters (converted to their corresponding star radius following an empirical calibration) affect the determination of the exoplanet radius.

\section{Analysis}

\subsection{Atmospheric parameters determination}

Stellar atmospheric parameters are determined studying absorption lines from spectroscopic observations. There are many different codes that can perform this kind of analysis, but we can say there are two major approaches: the equivalent width method (where only the area of the absorption lines are used, not their profile, and some assumptions are imposed such as excitation equilibrium and ionization balance) and the synthetic spectral fitting technique (where the observed spectra is compared to a synthetic one). Different spectroscopic analyzes can also vary in many other aspects such as the atomic data, model atmospheres, reference solar abundances, spectral resolution or the continuum normalization process. All these possible combinations can bias the results and do not play in favor of precision when we want to compare several stars analyzed in a heterogeneous way.

To study the impact of stellar atmospheric parameters in exoplanet characterization, I used stellar spectroscopic results derived with two set-ups where the spectroscopic technique is the only difference: equivalent width method and the synthetic spectral fitting technique. The analysis was done with iSpec\footnote{http://www.blancocuaresma.com/s/} \citep{2014A&A...569A.111B} using an atomic line list extracted from VALD \citep{2011BaltA..20..503K}, the MARCS\footnote{http://marcs.astro.uu.se/} model atmosphere \citep{2008A&A...486..951G}, solar abundances from \cite{2007SSRv..130..105G} and MOOG 2014 \citep{2012ascl.soft02009S} as the radiative transfer code. 

The analyzed high-resolution spectroscopic dataset \citep{2014A&A...566A..98B} corresponds to the Gaia FGK Benchmark Stars (GBS) \citep{2014A&A...564A.133J, 2015A&A...582A..81J, 2015A&A...582A..49H, 2016A&A...592A..70H}, a selection of very well-known stars accompanied with reference parameters. This collection is very convenient for this study since atmospheric parameters were determined independent of spectroscopy (except for their chemical composition), and the masses and radii were obtained by measuring the stellar diameter using interferometry and stellar models. More details about the spectroscopic analysis and its results can be found in \cite{2016csss.confE..22B} and \cite{2017hsa9.conf..334B}.

\subsection{Star selection}

\begin{figure}
\center
\includegraphics[width=\linewidth]{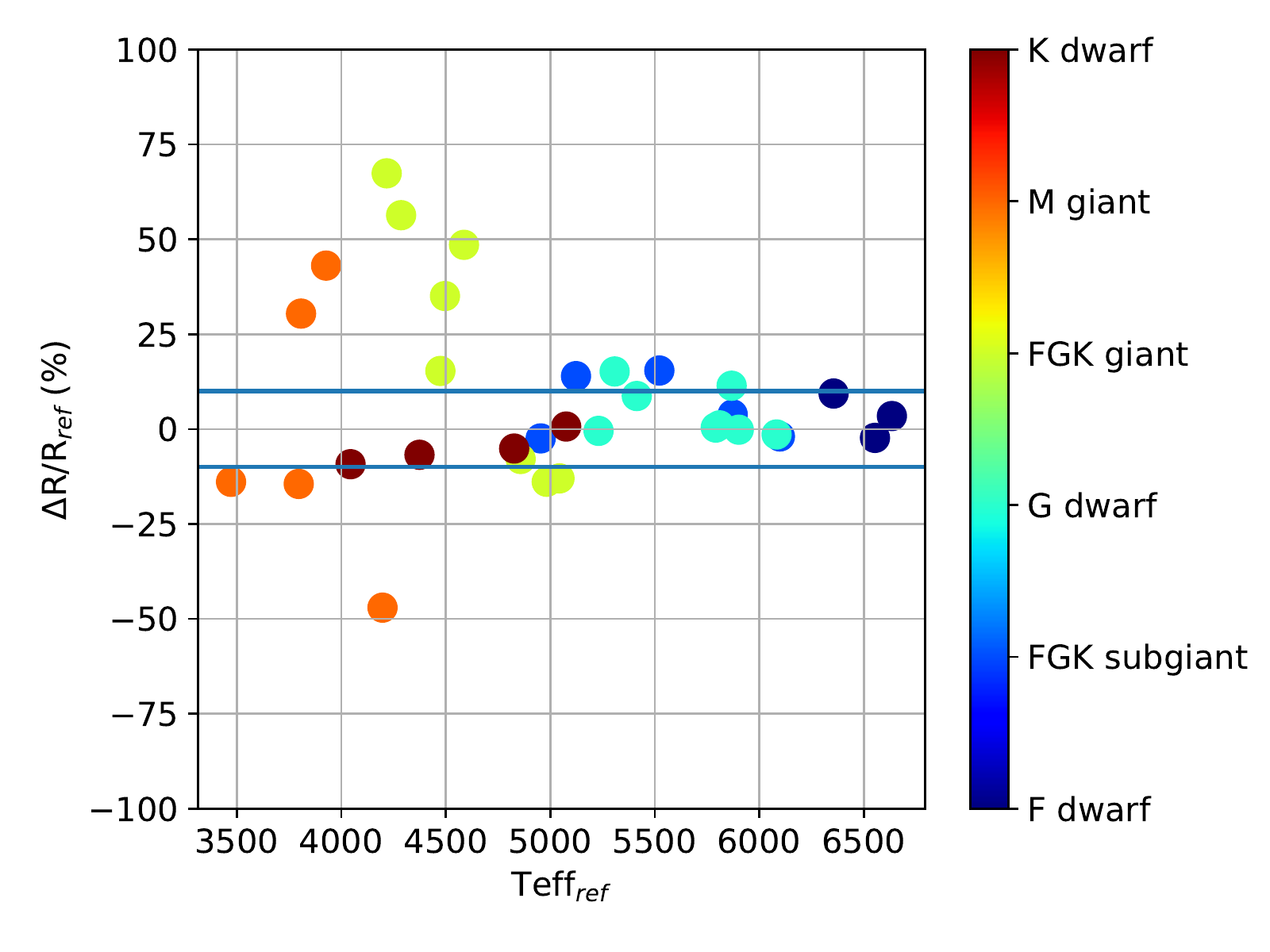}
\caption{\label{fig1} Difference between the estimated stellar radius (obtained using GBS reference atmospheric parameters and \cite{2010A&ARv..18...67T} relation) and the reference radius (in percentage). Horizontal lines included at $\pm10\%$ to indicate what GBS were accepted for the analysis.
}
\end{figure}

The empirical calibration from \cite{2010A&ARv..18...67T} provides a useful tool to derive stellar radius from atmospheric parameters. The authors used more than 90 detached non-interacting binary systems for which the mass and radius are known within errors of $\pm$3\%, and their stellar parameters were recomputed. With this data, they derived an empirical calibration for single stars above 0.6~M\textsubscript{\(\odot\)} which consists of two polynomial functions (page 110 and table 4 in their article) of effective temperature, surface gravity and iron abundance that yield stellar mass and radius. But as any empirical relation, its applicability is limited to a range of effective temperatures, surface gravities and metallicities. To evaluate which GBS can be safely used with this calibration, I used their reference atmospheric parameters to obtain stellar radius estimates, which then I compared to their reference radius as shown in Fig.~\ref{fig1}. 

The results are more accurate for dwarf/subgiant stars with metallicities close to the Sun, this was expected since most of the stars used to build the calibration cover this regime. For the next steps in the analysis, I discarded all GBS with a differential percentage of more/less than $\pm10\%$ of its reference radius in order to guarantee the correct use of the empirical relation.

\subsection{Stellar radius}

\begin{figure}
\center
\includegraphics[width=\linewidth]{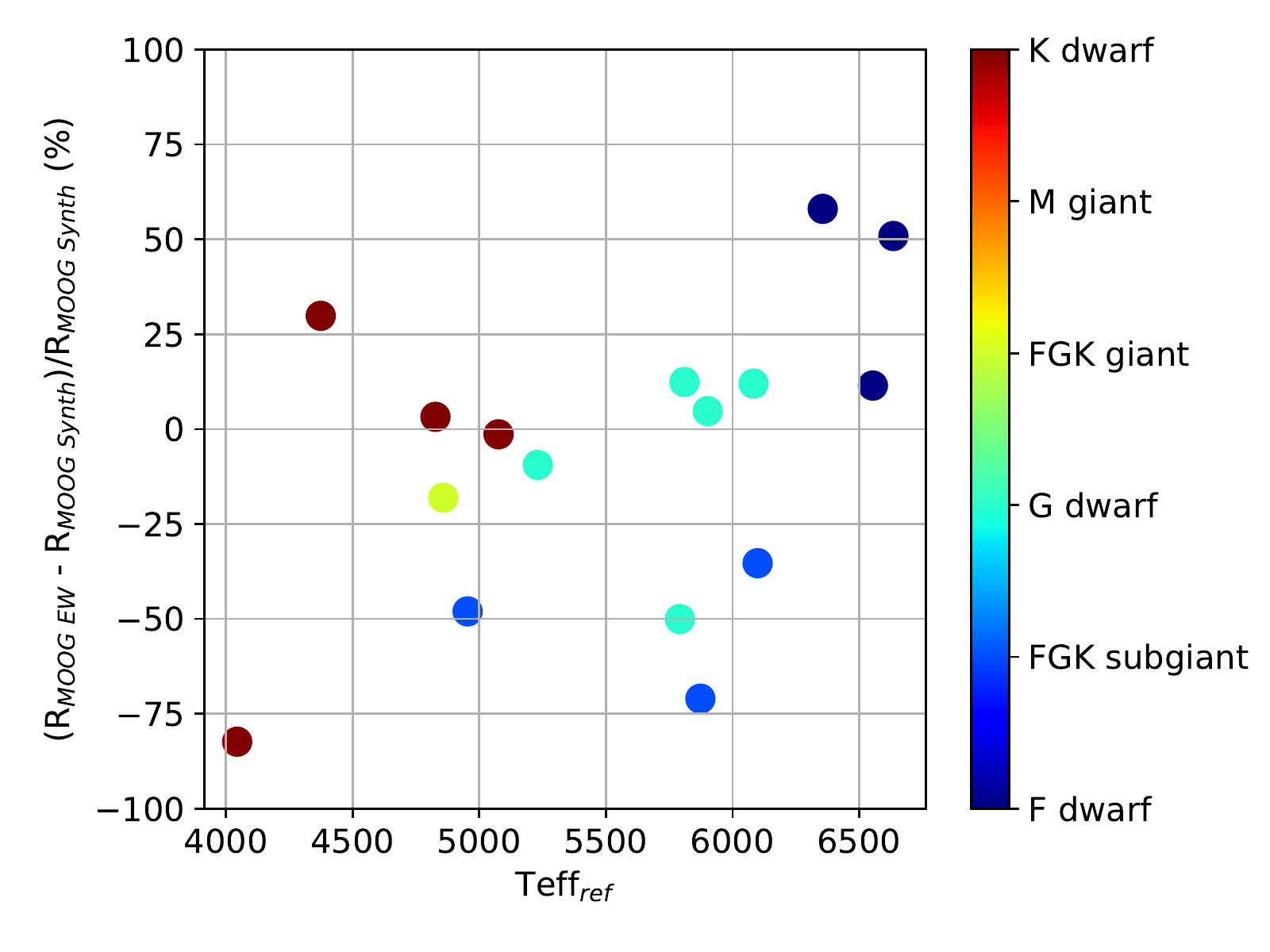}
\caption{\label{fig2} Difference between the estimated stellar radius obtained using atmospheric parameters obtained with the equivalent width method and the synthetic spectral fitting technique with MOOG as radiative transfer code (in percentage).
}
\end{figure}

In the second part of the analysis, I used the GBS atmospheric parameters derived with iSpec using the equivalent width method and the synthetic spectral fitting technique. Then I computed the corresponding stellar radius using the empirical calibration from \cite{2010A&ARv..18...67T} for each result and calculated the difference as shown in Fig.~\ref{fig2}.

The results show how the intrinsic differences between the two used spectroscopic techniques for the determination of atmospheric parameters can significantly impact the stellar radius derivation when using an empirical calibration. In the extreme cases, the estimated stellar radius with one method can be more than half of the radius derived with the other technique.

\section{Conclusions}

In this study I used atmospheric parameters derived for the Gaia FGK Benchmark Stars high-resolution spectral library using two approaches: the equivalent width method and the synthetic spectral fitting technique. After filtering the stars that cannot be used with the empirical calibration from \cite{2010A&ARv..18...67T}, I transformed the atmospheric parameters into stellar radius and compared how the differences in the former impact the latter. The results show that in extreme cases, the derived stellar radius can differ by more than 50\%. For stars with exoplanets, the exoplanet radius can be derived by using the radius ratio (obtained from transit observations) and the stellar radius (Eq.~\ref{eq:transit}). Hence, the stellar discrepancies propagate directly to the estimation of the exoplanet radius.

Exoplanet radiuses are used to infer its possible composition and to distinguish between rocky and gaseous planets \citep{2016ApJ...819..127Z}, bad stellar radius estimates are going to have a relevant impact on the characterization of the planetary system. For instance, a 50\% difference in planetary radius implies that, for a given planetary mass, an exoplanet could either be rocky with a 25\% iron composition or gaseous with more than 50\% of water composition\footnote{https://www.cfa.harvard.edu/\textasciitilde lzeng/planetmodels.html}. Usually the radius of the planet is one of the planetary parameters given with the highest precision (down to a few percents), but I show here that the radius determination of the star can be critical. Thus, it is extremely important to execute homogeneous spectroscopic analysis and use well-calibrated stellar results to be able to compare planetary systems in a trustworthy manner.

\section*{Acknowledgments}
{This work would not have been possible without the support of Dr. Laurent Eyer (University of Geneva). This research has made use of NASA's Astrophysics Data System.}

\bibliographystyle{ewass_ss4proc}
\bibliography{stellar_parameters_and_exoplanets.bib}

\end{document}